# Spin-wave Resonance in Arrays of Nanoscale Synthetic-antiferromagnets

Vladyslav Borynskyi
*Department of Physics of Films*
*Institute of Magnetism of the NAS of Ukraine and MES of Ukraine*
Kyiv, Ukraine
vladislav.borinskiy@gmail.com

Anatolii Kravets
*Department of Physics of Films*
*Institute of Magnetism of the NAS of Ukraine and MES of Ukraine*
Kyiv, Ukraine
anatolii@kth.se

Dmytro Polishchuk
*Nanostructure Physics*
*KTH Royal Institute of Technology*
Stockholm, Sweden
dpol@kth.se

Alexandr Tovstolytkin
*Department of Physics of Films*
*Institute of Magnetism of the NAS of Ukraine and MES of Ukraine*
Kyiv, Ukraine
atov@imag.kiev.ua

Iryna Sharai
*Department of Physics of Films*
*Institute of Magnetism of the NAS of Ukraine and MES of Ukraine*
Kyiv, Ukraine
ira-sharay@ukr.net

Vladislav Korenivski
*Nanostructure Physics*
*KTH Royal Institute of Technology*
Stockholm, Sweden
vk@kth.se

Andrii Melnyk
*Center for the collective use of scientific equipment "EPR spectroscopy"*
*Institute for Sorption and Problems of Endoecology of the NAS of Ukraine*
Kyiv, Ukraine
mak106@ukr.net

*Abstract*—The study concerns dynamics of standing spin waves in arrays of sub-100 nm elliptic synthetic-antiferromagnet (SAF) nanodisks. We performed a detailed ferromagnetic resonance analysis in conjunction with micromagnetic modeling to find out several prominent traits of such systems. One broad line is shown to be the sole resonant response for a SAF of the considered sizes. We demonstrate that this mode is degenerated, and its excitation map resembles a superposition of in-center and edge-type oscillations. We also show how this hybrid excitation leads to almost twofold enhancement in the shape-induced anisotropy of the mode.

## I. INTRODUCTION

Continuous advances in fabrication methods of small nanoelements, with lateral sizes down to tens of nanometer, brought technological capabilities for rapid development in fields of spintronics and magnonics. Extensive research into magnetic nanostructures has been progressing for the last decades, providing new functionalities in modern devices such as neuromorphic junctions, spin-torque oscillators as well as magnonic crystals [1], [2]. As the element sizes are decreased to the scale of characteristic exchange lengths, e.g. around 5 nm for Py (permalloy, $Ni_{80}Fe_{20}$), the observed magnetization dynamics transforms significantly. Standing spin waves, which are driven by dipole interaction and typically reside in high-frequency range, disappear, and only the well-known center- (uniform) and edge-mode (localized at the element edges) oscillations persist [3]. This is the result of the change in the coupling type: magnetostatic energy contribution, dominating in bigger magnets, is surpassed by direct exchange in smaller ones within an individual layer.

Magnetic behavior of nanostructured synthetic antiferromagnets (SAFs) or ferrimagnets is also determined by the type of dominating interlayer coupling [4]–[6]. Such systems generally consist of multiple ferromagnetic (FM) layers spaced by nonmagnetic (NM) or paramagnetic interlayers. Depending on the spacer composition and its magnetic state, the outer FM layers can be magnetically aligned either parallel or antiparallel [7]–[10]. This internal design of SAF nanostructures allows engineering their magnetic ground states and, importantly, new mechanisms for magnetization switching [11]. Apparently, this approach is effective for tailoring the spin-wave dynamics of such SAF systems. To note, upon integrating these nanostructures into devices, consecutive difficulties might be faced e.g. fabrication flaws or shape roughness [12]–[14]. The suggested problems, along with such an abundant functionality of SAF-based systems, demand comprehensive characterization in the on-going research.

Recent studies have shown that SAF nanodisks exhibit additional higher-order resonance modes [15]–[17], which can be modified by the symmetry of arrays, number of layers and their thicknesses as well as geometric parameters of the elements. Recently we have demonstrated how the specific interaction inside the SAF allow direct control on spin-wave resonance in nanoarrays, introducing ways for tuning the high-speed operation of possible spintronic applications [17]. Here, we demonstrate that the reported dynamics undergo substantial changes with changing the size and shape of SAF nanodisks.

## II. METHODS

Multilayers Py(7 nm)/NiCu(10 nm)/Py(7 nm), assembling the initial film sample, were deposited onto a silicon substrate by Ar sputtering (Orion, AJA Int.) with TaN serving as a protective cover. The 20% content of copper in the doped NiCu alloy guaranteed antiparallel i.e., antiferromagnetic ground state of SAF [7], [10]. The array was afterwards fabricated using electron-beam lithography (Voyager, Raith Inc.), where the subsequent process of plasma etching ensured an elliptical shape for each SAF element. The use of ferromagnetic resonance (FMR) X-band spectrometer (ELEXSYS E500, Bruker Inc.) granted cavity-averaged response at constant microwave frequency 9.35 GHz. Supporting micromagnetic simulations, performed in MuMax [18], [19], provided respective calculated spectra and enabled spatial spin-wave characterization, consistent with experimental data. During the simulations, the layout of SAF elements, the inter-element spacing as well as material parameters remained strictly close to those for the fabricated array.

## III. RESULTS

### A. A single hybrid mode

The array geometry and the experimental setup can be seen from the scanning electron microscopy (SEM) image in Fig. 1(a). An average length for the major axis of an arbitrary SAF element in the array is 75 nm, whereas the minor axis is about 0.7 of that value. We measured a series of cavity-FMR spectra, each corresponding to a specific azimuthal orientation $\varphi_H$ of the external magnetic field. Two spectra, selected for angles 0 and 90°, are depicted in Panel (b). All measured spectra reveal one pronounced resonance line, located above the resonance field of a conventional Py film. Importantly, both positions of the single resonance line, at 0 and 90°, are significantly higher than the expected SAF saturation field [10], [20]. Thus, vortex, spin-flop or other exotic magnetic states [21]–[23] cannot be the source of this signal.

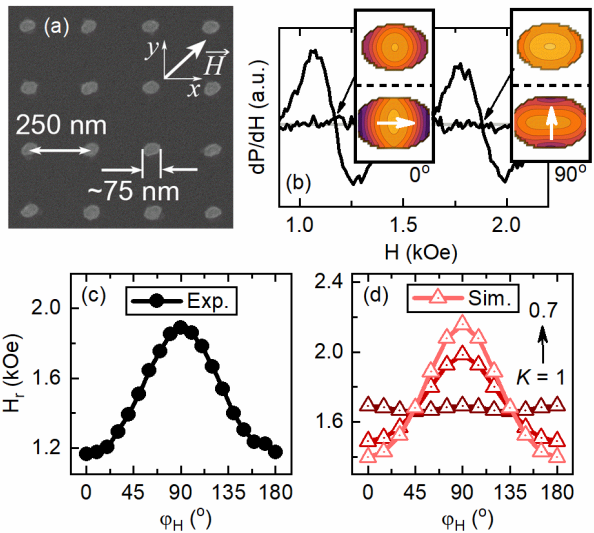

Fig. 1. (a) SEM image of the fabricated array. (b) Experimental FMR spectra measured for the array at 0 and 90° directions of the external field. Insets show corresponding excitation maps for each layer of the SAF. (c-d) Angle-dependent resonance field of the mode obtained from experimental and calculated data, respectively. Symbol $K$ denotes minor-to-major axis ratio of the SAF nanodisk.

Excitation maps, depicted in the insets in Fig. 1(b), show spatial intensity distribution of oscillations and evidence the spin-wave origin of the resonance. Our previous study demonstrated three-mode dynamics in 150-nm nanodisks [17], where dipole interactions within the SAF promoted splitting of the edge-mode into acoustic and quasi-optical regimes. Here, however, only the single degenerate mode is excited, what we explain by the dominance of the exchange energy over magnetostatics in small elements [3]. The spin excitation maps indicate another important consequence that the observed single mode cannot be solely ascribed to neither center, nor edge-type resonance, and is hybrid between the two.

### B. An enhancement of shape anisotropy

The angular dependencies of the resonance field, presented in Fig. 1(c-d), indicate a substantial shift of the resonance line by ~700 Oe when changing orientation of the external magnetic field from 0 to 90°. The ellipticity of SAF elements, clearly visible in Panel (a), is one of the reasons behind this behavior. Indeed, $H_r$ dependencies, calculated for a set of ellipticity values $K$ (minor-to-major axis ratio) and depicted in Panel (d), clearly show the transition from weak bi-axial anisotropy, common for such arrays [24], [25], to uniaxial one, dictated by the shape anisotropy of an individual element. However, analytical estimations for SAF elements of the considered size provide significantly smaller value of the respective anisotropy field. The difference in the demagnetizing field from major- to minor-axis SAF saturation reaches as high as 450 Oe [26]. We suggest that not only element shape is responsible for the revealed strong anisotropy, but also the specific configuration of the hybrid resonance mode, as its spatial distribution is evidently unequal between 0 and 90° magnetization directions.

## IV. DISCUSSION

Our results demonstrate substantial modifications of magnetization dynamics in sub-100 nm SAF nanodisks, arguably governed by the interplay between the shape anisotropy and the single hybridized mode. The oscillation type of this mode resembles the combination of the well-known center and edge modes, and draws several important conclusions discussed below. Firstly, the presence of excitations throughout the SAF nanostructure, both in the middle and border regions, makes this mode less sensitive to morphological deviations of elements, decreasing possible spread of the resonance frequency [13]. Moreover, SAF-induced dipole coupling is surpassed by direct exchange within an individual layer, which effectively flattens the excitation across the disk. As a result, we observe the single broad resonance line ($\Delta H_{av}$ = 300 Oe). Secondly, despite the fact that oscillations embrace the entire SAF area, they are not completely uniform. Notably higher amplitude is seen at the edges of each layer. Under these circumstances, the ellipticity of elements causes difference in edge-to-edge mode coupling, complementing the effect of static demagnetization and enhancing the uniaxial shape anisotropy. Thus, higher energy is required to excite the resonance along the minor axis direction. These findings suggest new perspectives for integration of SAF-based structures in modern spintronic devices and should be relevant in further nanomaterial applications.

ACKNOWLEDGMENT

Support from the National Academy of Sciences of Ukraine (0122U002260 and 0122U001885), the Swedish Research Council (VR 2018-03526), the Olle Engkvist Foundation (project 2020-207-0460), and the Volkswagen Foundation (Grant No. 97758) are gratefully acknowledged.